\documentclass[acmsmall,nonacm]{acmart}

\title{Guiding user annotations for units-of-measure verification}
\usepackage{fancyvrb}
\usepackage{wrapfig}
\definecolor{Maroon}{rgb}{0.5,0.2,0}

\author{Dominic Orchard}
\affiliation{
  \institution{University of Kent}
}
\email{D.A.Orchard@kent.ac.uk}
\orcid{0000-0002-7058-7842}

\author{Mistral Contrastin}
\affiliation{
  \institution{Facebook London}
}

\author{Matthew Danish}
\affiliation{
  \institution{University of Cambridge} %% \institution is required
}
\email{mrd45@cam.ac.uk}          %% \email is recommended

\author{Andrew Rice}
\affiliation{
  \position{Reader}
  \institution{University of Cambridge} %% \institution is required
}
\email{acr31@cam.ac.uk}          %% \email is recommended

\begin{abstract}
  This \textbf{extended abstract}\footnote{Presented at HATRA 2020
  (Workshop on Human Aspects of Types and Reasoning Assistants)
  colocated with SPLASH 2020.} reports on previous work of the CamFort
project in which we developed an external units-of-measure type system
for Fortran code, targeted at scientists. Our approach can guide the
programmer in adding specifications (type annotations) to existing
code, with the aim of easing adoption on legacy code.
Pertinent to the topics of the HATRA workshop, we discuss the human-aspects
of the tool here.
CamFort is
open-source and freely available
online.\footnote{\url{http://camfort.github.io/}}
\end{abstract}

\begin{document}
\maketitle

\noindent
In modern science, it is common for models of the real
world to be expressed as computer programs. Such programs are often
highly numerical, dealing in quantities expressing real-world values.
Most mainstream programming languages can only treat such values as
uniformly typed (usually floats or doubles), yet the meaning of these
values is often much richer: they have a \emph{dimension} (e.g., time or
length) and a \emph{unit-of-measure} (e.g., seconds or metres), and perhaps even a \emph{kind of
  quantity}~\cite{foster2017quantity} (e.g., angular momentum or
linear momentum). Checking the consistency of units-of-measure or
dimensionality in equations is a
method long employed by scientists~\citep{Macagno:1971}. However, units checking has not been
widely adopted in scientific programming despite the familiarity of the
technique and a multitude of proposed systems for verification of
units-of-measure~\citep{Ore:2017:ISSTA,Contrastin:2016,gmpreussner,Jiang:2006:ICSE,Kennedy:2009:CEFP,Snyder:2013:N1969}
(many influenced by the work of~\citet{Kennedy:1996} on expressing
units-of-measure as a type system). One explanation for this lack of
adoption is the burden of adding units annotations to existing
codebases, which may be large~\cite{salah2019understanding}. For
example, we previously observed a ratio of $\approx$1:10 between
variable declarations and lines of physical code in scientific Fortran
code~\cite{Orchard:2015:JCS}. If each variable requires a
units-of-measure specification then the burden of adopting such a
system becomes prohibitively large for even medium-sized code bases
(we found 10kloc programs had about 1k
variables~\cite{Orchard:2015:JCS}). For code bases with over a
million lines (e.g., the Met Office's Unified Model for
climate and weather forecasting~\cite{um}) writing
$\approx$100,000 specifications is even more of a non-starter.

However, given an adequate type inference procedure, it is almost
certainly not necessary for a programmer to annotate every variable
with its unit. For example, for \texttt{v = d / t}, we need only know
the units of \texttt{d} and \texttt{t} to determine \texttt{v}
automatically. Indeed, in a study of a corpus of small Fortran
programs, we found that unit-of-measure inference could reduce the
annotation burden by about 80\%~\cite{Orchard:2015:JCS}. Even with
this reduction, how does a programmer
faced with perhaps thousands of variables decide which ones need
a type annotation? In the above example it takes a moment's
thought to realise that annotating just \texttt{v} is not adequate to
determine the units of \texttt{d} and \texttt{t}. Thinking about
such constraints over an entire program quickly becomes infeasible
for a programmer.

Yet units-of-measure typing could
be highly beneficial in science (and engineering) where the
stakes of making an error are high (the famous example
is the loss of the \$327 million Mars Climate
Orbiter
due to a mismatch of Imperial vs. metric
units~\citep{Stephenson:1999:MCO}). Our goal was to develop
a tool addressing the human aspects of units-of-measure
verification to overcome these barriers to entry.

\subsection*{Guiding the user -- suggesting which variables to annotate}

CamFort is a suite of verification tools, one
of which provides units-of-measure types to Fortran as
\emph{extrinsic} properties of values (in the sense
of~\citet{reynolds2000meaning}) expressed as comments and
checked or inferred by the external tool. The approach allows annotations to be
incrementally added to a code base~\cite{unitsnew}, differing
to F\#'s units-of-measure typing which are not easily applied
incrementally.

CamFort provides a `suggest' feature
to address the problem of how a user knows which variables
to annotate to maximally exploit inference
and reduce the annotation burden.
This feature reports the
\emph{critical variables}: a minimal (possibly non-unique) subset of program variables for which knowing
their units determines the units of the remaining variables
not in the critical set. This set is computed by applying
the usual constraint generation for inference/checking
of the units types. Such constraints are expressed
as a matrix, on which Gaussian elimination is applied to compute
a solution. For
an under-constrained system, the variables
which require an annotation to yield an overall solution
can be identified from the resulting
reduced
row-echelon form of the matrix~\cite{Orchard:2015:JCS}.

\begin{wrapfigure}{r}{0.22\linewidth}
\vspace{-1em}
\begin{Verbatim}[fontsize=\footnotesize,numbers=left,xleftmargin=0em,numbersep=-0.2em,commandchars=\\\{\}]
  {\textcolor{blue}{real}} :: a, b
  {\textcolor{blue}{real}} :: x = 20.0
  {\textcolor{blue}{real}} :: t = 3.0
  a = sqr(x)
  b = sqr(t)

  {\textcolor{gray}{contains}}
  {\textcolor{blue}{real}} {\textcolor{teal}{function}} sqr(y)
    {\textcolor{blue}{real}} :: y
    sqr = y * y
  {\textcolor{teal}{end}} {\textcolor{teal}{function}}
\end{Verbatim}
\vspace{-0.7em}
\caption{Sample code}
\label{fig:sample}
\vspace{-4em}
\end{wrapfigure}

As a demonstration,
consider the (slightly abridged) Fortran program in Figure~\ref{fig:sample}. Running
\texttt{`camfort units-suggest'} on this code returns:
\begin{Verbatim}[fontsize=\footnotesize]
sample.f90: 2 variable declarations suggested to be given a specification:
    sample.f90 (3:11)    t
    sample.f90 (2:11)    x
\end{Verbatim}
The user can specify their units-of-measure via type
annotation comments written above the variable declarations, e.g., adding
{\textcolor{Maroon}{\texttt{!= unit(m) :: x}}} and {\textcolor{Maroon}{\texttt{!= unit(s) :: t}}} before lines 2
and 3 respectively. The user can then use the \texttt{units-infer} mode
to see the inferred unit specifications for the rest of the code,
or \texttt{units-check} to check the code against the specifications.

\begin{wrapfigure}{l}{0.37\linewidth}
\vspace{-1em}
\begin{Verbatim}[fontsize=\footnotesize,numbers=left,xleftmargin=1em,numbersep=0em,commandchars=\\\{\}]
  {\textcolor{Maroon}{!= unit(m**2) :: a}}
  {\textcolor{Maroon}{!= unit(s**2) :: b}}
  {\textcolor{blue}{real}} :: a, b
  {\textcolor{Maroon}{!= unit(m) :: x}}
  {\textcolor{blue}{real}} :: x = 20.0
  {\textcolor{Maroon}{!= unit(s) :: t}}
  {\textcolor{blue}{real}} :: t = 3.0
  a = sqr(x)
  b = sqr(t)

  {\textcolor{gray}{contains}}
  {\textcolor{Maroon}{!= unit(('a)**2) :: sqr}}
  {\textcolor{blue}{real}} {\textcolor{teal}{function}} sqr(y)
    {\textcolor{Maroon}{!= unit('a) :: y}}
    {\textcolor{blue}{real}} :: y
    sqr = y * y
  {\textcolor{teal}{end}} {\textcolor{teal}{function}}
\end{Verbatim}
\vspace{-0.7em}
\caption{After spec. synthesis}
\label{fig:synth}
\vspace{-2.2em}
\end{wrapfigure}

\subsection*{Synthesising types and inserting them into users' source code}

Another useful feature of the tool is that it
can synthesise any inferred specifications into the user's
code for them, inserting the inferred type annotations at relevant
points. For example, following \texttt{units-suggest} and the
user giving annotations to \texttt{x} and \texttt{t}, the user
can run \texttt{`camfort units-synth'} which runs inference
and then inserts any specifications not already
present into the source code. The result for the
Figure~\ref{fig:sample} program is shown in Figure~\ref{fig:synth}. Note
that the \texttt{sqr} function is unit-polymorphic and so is
given a polymorphic specification, where \texttt{'a}
means any unit (using ML-style syntax).

This specification synthesis technique is provided by other
verification features of
CamFort~\cite{DBLP:journals/pacmpl/OrchardCDR17,contrastin2016supporting}.
We think this is an underused technique: compilers with
type inference could potentially provide an analogous feature, aiding user understanding
and documentation.

\subsection*{Lessons learned}

We think there are two useful ideas here that can be applied elsewhere
to improve human-computer interaction with verification tools: (1) use
the constraints of checking or inference to guide the user to insert
a minimal set of specifications which maximise information; (2) synthesise inferred
specifications into source code. This may be best done `on request',
at specific locations given by the user, akin to how program synthesis
tools can insert synthesised programs directly into code. A similar
idea is seen in type qualifier inference tools such as
Cascade~\cite{vakilian2015cascade}, with guided annotation and
specification generation.

\bibliography{references}

\end{document}